\documentclass[11pt]{article}
\usepackage{amsmath,amssymb,color,graphics,epsfig,cite}
\usepackage{amsfonts}
\newcommand{\hoch}[1]{$\, ^{#1}$}

\makeatletter
\@addtoreset{equation}{section}
\makeatother

\newcommand{\ba}{\begin{eqnarray}}
\newcommand{\ea}{\end{eqnarray}}

\def\ben{\begin{equation}}
\def\een{\end{equation}}
\def\bea{\begin{eqnarray}}
\def\eea{\end{eqnarray}}
\def\be{\begin{equation}}
\def\ee{\end{equation}}

\def\nn{\nonumber}

\def\ben{\begin{equation}}
\def\een{\end{equation}}
\def\bea{\begin{eqnarray}}
\def\eea{\end{eqnarray}}
\def \nn {\nonumber}

\def\ft#1#2{{\textstyle{\frac{\scriptstyle #1}{\scriptstyle #2} } }}
\def\fft#1#2{{\frac{#1}{#2}}}

\begin{document}

%\begin{flushright}
%\hfill{MI-TH-1533}
%\end{flushright}

%\vspace{25pt}
\begin{center}
%{\Large {\bf Butterfly Velocity, Thermodynamical Volume, \\ Reverse Isoperimetric Inequality and the Velocity Bound}}

{\Large {\bf On the Size of Rotating Black Holes}}

\vspace{20pt}
{ Xing-Hui Feng and  H. L\"u\hoch{*}}

\vspace{10pt}

{\it Center for Joint Quantum Studies and Department of Physics,\\
School of Science, Tianjin University, Tianjin 300350, China}

\vspace{40pt}

\underline{ABSTRACT}
\end{center}

Recently a sequence of inequalities relating the black hole horizon, photon sphere, shadow were proposed for spherically symmetric and static black holes, providing the upper bound for given mass.  In this paper, we extend the discussion to include rotating black holes. When viewed from the north pole direction, the shadow remains a round disk, but the image is skewed when viewed from the equatorial plane.  After properly implementing the ``size'' parameters for the rotating black holes, we verify that the sequence of inequalities remain valid for a variety of solutions, including Kerr, Kerr-Newman, Kerr-Sen and Kerr-Cveti\v c-Youm black holes. The upshot is that rotation makes both the actual and apparent sizes of a black hole smaller.

\vfill {\footnotesize xhfeng@tju.edu.cn, \quad mrhonglu@gmail.com}

{\footnotesize \hoch{*}Corresponding author}

\thispagestyle{empty}

\pagebreak

\tableofcontents
\addtocontents{toc}{\protect\setcounter{tocdepth}{2}}

%%%%%%%%%%%%%%%%%%%%%%%%%%%%%%%%%%%%%%%%

\newpage
%%%%%%%%%%%%%%%%%%%%%%%%%%%%%%%%%%%%%%%%

\section{Introduction}

Since Sir Author Eddington's eclipse experiment one hundred years ago, there have been continuing efforts in understanding and observing the null geodesics around gravitationally intense massive objects including black holes. Since the earlier work on the null geodesics of the Schwarzschild black hole \cite{Synge:1966okc,Luminet:1979nyg}, a general picture emerges.  Massless particles like photons can form an unstable surface outside the the black hole event horizon. Photons inside whose orbits do not cross the surface will be trapped,\footnote{Trapped photons are generally expected to spiral into the black hole horizon; however, recently new black holes satisfying the dominant energy condition was constructed \cite{Liu:2019rib}, for which trapped photons could also form a stable photon shield outside the black hole horizon \cite{Liu:2019rib,Lu:2019zxb}.} whilst those outside whose orbits do not cross the surface will escape to infinity, surrounding a shadow of the black hole [3-39]. This area of research was further boosted recently by an actual photo of such a shadow \cite{Akiyama:2019cqa}.

The subject is significantly simpler for spherically symmetric black holes, for which the horizons, photon spheres and shadow disks are all round, whose sizes can be characterized by their radii. The simplicity allows one to establish some universal relations among these geometric surfaces.  Based on the dominant energy condition, together with the negative trace of the energy momentum tensor, Hod proved an upper bound for photon sphere radius $R_{\rm ph}$ for a black hole of mass $M$, namely \cite{Hod:2017xkz}
\be
R_{\rm ph}\le 3M\,.
\ee
Based on the same energy condition, Cveti\v c, Gibbons and Pope found an inequality relation between the radii of photon sphere and the shadow disk \cite{Cvetic:2016bxi}
\be
R_{\rm ph}\le \fft{R_{\rm sh}}{\sqrt3}\,.
\ee
While the radii $R_{\rm ph}$ and $R_{\rm sh}$ of the photon sphere and shadow disk measure the apparent sizes of a spherically symmetric black hole, the actual size is determined by the horizon radius $R_+$.  Penrose's conjecture \cite{Penrose:1973um} of the black hole entropy upper bound is equivalent to the Riemannian Penrose inequality
\be
R_+\le 2M\,,
\ee
which is largely considered as been proven under the dominant energy condition, see e.g.~\cite{Mars:2009cj}.
Recently a sequence of inequalities relating all these size parameters of spherically symmetric black holes was proposed, namely \cite{Lu:2019zxb}
\be
\fft32 R_+\,\le R_{\rm ph}\, \le\, \fft{R_{\rm sh}}{\sqrt3}\,\le\, 3 M\,.\label{conjecture}
\ee
In \cite{Lu:2019zxb}, many different types of black holes satisfying at least the null energy condition were examined to verify this set of inequalities and no counterexample was found.

However, a gravitationally intense massive object or a black hole in our Universe, formed from gravitational collapse, typically have large angular momentum and they are not spherically symmetric, but axially symmetric.  We shall use the mean radius $R_+=\sqrt{{\cal A}_+/(4\pi)}$ to characterize the horizon size, where ${\cal A}_+$ is the area of the horizon.  This definition of $R_+$ continues to be relevant in Penrose's entropy conjecture for rotating black holes.

The size of the black hole shadow is more subtle to characterize. The shape changes depending on the latitude angle $\theta_0$ of the observer at the asymptotic infinity, with respect to the north pole direction of the rotating black hole. When viewed from the north pole $(\theta_0=0)$, the shadow is round regardless of the angular momentum.  The shape will be skewed as we increase the angle $\theta_0$ and become most distorted at the equatorial plane ($\theta_0=\pi/2$).  It is thus tricky to find a parameter to characterize the size of such a shadow.  One possibility is to consider the area of the shadow, which we present some results in appendix \ref{sec:app-shadow}.  In this paper, we shall focus on a simpler alternative measure.
As we can see from the shadow graphs in appendix \ref{sec:app-shadow}, the boundary of the shadow is a closed convex loop. For any given point in the loop, there exists a longest (or more precisely local extremal) diagonal line joining a point at the other side of the loop. Among these diagonal lines, we find that there exists a shortest one and we define $R_{\rm sh}$ as its half length.  In general $R_{\rm sh}$ depends on the viewing angle $\theta_0$.  When $\theta_0=0$, the shadow is round and $R_{\rm sh}$ becomes simply the radius of the disk.

The exact shape of the photon surfaces is also hard to determine in general. In fact they form a thick region instead of a thin shell when the black hole is rotating \cite{Grenzebach:2014fha}.  However, we note that the photons appearing in the boundary of the shadows come from some specific photon orbits that depend also on the viewing angle $\theta_0$.  We thus consider a definition of $R_{\rm ph}$ based on $g_{\theta\theta}(r_{\rm ph}, \theta_0)$, where $r_{\rm ph}$ is the radial coordinate of the relevant photon orbit.

The purpose of this paper is to establish $R_{\rm ph}$ and $R_{\rm sh}$ for the general metric ansatz of rotating black holes and verify the validity of the conjecture (\ref{conjecture}) using explicit examples of black holes.  The paper is organized as follows. In section 2, we set up the general formalism to compute the photon orbits and resulting photon shadows.  We consider two classes of metric Ans\"atze for rotating black holes and obtain their geodesic motions.  We obtain the formulae for both photon orbits and black hole shadows. We then argue and present the definitions of $R_{\rm ph}$ and $R_{\rm sh}$ that are dependent on not only the black hole parameters, but also the viewing angle $\theta_0$ of the asymptotic observer.  In sections 3,4,5,6, we consider Kerr, Kerr-Newman, Kerr-Sen and Kerr-Cveri\v c-Youm black holes and verify the validity of the conjecture (\ref{conjecture}). We conclude the paper in section 7.  In appendix \ref{sec:app-shadow}, we present a few numerical plots of Kerr black hole shadows, which help to illustrate the logic of our definitions of the size parameters.  In appendix \ref{sec:quartic}, we give the solutions to the general quartic polynomial equation, which appears in a few black hole examples in determining the photon orbits.

\section{Photon orbits and black hole shadows}
\label{sec:photon}

\subsection{Photon orbits}

Unlike spherically symmetric and static black holes, the metric ansatz for the stationary rotating black holes is much more complicated.  For our purpose, it is clearly advantageous to consider analytical solutions.  Even if two metric Ans\"atze are equivalent via coordinate transformation, they may not be both analytical in closed form. On the other hand, it can be tedious to discuss the null geodesic motion repetitively  for each black hole solution.  In this section, we present two classes of metric Ans\"atze for rotating black holes. Both are in the Boyer-Linquiste coordinates, asymptotically to the flat {\it non-rotating} Minkoski spacetime with
\be
ds^2 =- dt^2 + dr^2 + r^2 \big( d\theta^2 + \sin^2\theta\, d\phi^2\big).
\ee
The form of the class-one metrics was already in literature.  We introduce the class-two metric to cope with more general solutions that cannot be cast into the class-one metric ansatz.

It is worth mentioning that rotating black holes, with axial symmetry, are of cohomogeneity two, depending on the radial $r$ and latitude angle $\theta$ only. We choose a coordinate gauge where the cross terms in the metric involve only the time $t$ and the longitudinal angle $\phi$.  The constant time and radial slices are two-dimensional spheres that are generally not round.  In this paper, the quantity $g_{\theta\theta}(r,\theta)$ plays an important r\^ole in deciding the size of the orbits.

\subsubsection{Class-one metric ansatz}

The class-one metric ansatz was introduced in \cite{Shaikh:2019fpu}, and it takes the form
\bea
ds^{2}& =&-F d t^{2}-2 a \sin ^{2} \theta\,\left(\sqrt{\ft{F}{G}}-F\right) d t d \phi+\frac{H}{G H+a^{2} \sin ^{2} \theta} d r^{2}+H d \theta^{2} \nn\\
&&+\sin ^{2} \theta\left(H+a^{2} \sin ^{2} \theta(2 \sqrt{\ft{F}{G}}-F)\right) d \phi^{2}.\label{classone}
\eea
The functions $F,G$ and $H$ depend on both the radial $r$ and latitude $\theta$.  However, they are subject to the constraints
\bea
\Delta(r) &=& G(r, \theta) H(r, \theta)+a^2 \sin^2\theta \,,\nn\\
X(r) &=&\sqrt{\ft{G(r, \theta)}{F(r, \theta)}} H(r, \theta)+a^{2} \sin^{2} \theta\,.\label{DeltaX}
\eea
In other words, the specific combinations $(\Delta, X)$ are functions of $r$ only.  One disadvantage of this metric ansatz, however, is that without additional input of the metric functions, the horizon geometry cannot be abstractly analysed.  In fact, we have $F=G$ in all the explicit examples we present in this paper that can be written in the form of (\ref{classone}).  In these examples, the metrics are thus specified completely by the $r$-dependent functions $(\Delta, X)$, the functions $F=G$, $H$ and hence the whole metric can be determined from (\ref{DeltaX}).  In this $F=G$ case, the entropy of the black hole is simply given by
\be
S=\pi X(r_+)\,,\label{entropy1}
\ee
where $r_+$ is the largest root of $\Delta(r)$. It follows from the discussion in the introduction, the mean radius of the horizon is
\be
R_+=\sqrt{X(r_+)}\,.
\ee
It is curious to note that we have the following equality and inequality relations
\be
S=\pi g_{\theta\theta}(r=r_+, \theta=0) \ge \pi g_{\theta\theta}(r_+,\theta)\,.\label{sGthetatheta}
\ee
We shall comment on the significance of this inequality later.

The null geodesic equations associated with the class one metric (\ref{classone}) can be obtained from the Hamilton-Jacobi equation. These give rise to three integration constants and four first-order differential equations
\bea
\frac{F}{G} \Delta(r) \frac{d t}{d\tau}&=& E \left(H+a^{2} \sin ^{2} \theta(2 \sqrt{\ft{F}{G}}-F)\right)-a L\left(\sqrt{\ft{F}{G}}-F\right),\nn \\
\frac{F}{G} \Delta(r) \frac{d \phi}{d\tau}&=& E a\left(\sqrt{\ft{F}{G}}-F\right)+\fft{L F}{\sin ^{2} \theta}\,,\nn\\
H \frac{d r}{d\tau}&=&\pm \sqrt{{\cal R}(r)}\,,\nn\\
H \frac{d \theta}{d\tau}&=&\pm \sqrt{\Theta(\theta)}\,,\label{geodesic1}
\eea
where $\tau$ is the affine parameter and
\bea
{\cal R}(r)&=&(E X-a L)^2-\Delta\,(Q+(L-aE)^2)\,,\nn\\
\Theta(\theta)&=&Q+E^2a^2\cos^2\theta-L^2\cot^2\theta\,.
\eea
Here $(E, L, Q>0)$ are three integration constants specifying the null geodesics. It is of interest to note that unless the rotating parameter $a$ vanishes, there can be no orbit with constant longitudinal angle $\phi$. Orbits with constant latitude angle $\theta$ is possible with appropriate integration constant $Q$.

The radial location $r_{\rm ph}$ of the unstable photon orbit is determined by
\be
R(r_{\rm ph})=0\,, \qquad R'(r_{\rm ph})=0\,, \qquad R''(r_{\rm ph})\ge 0\,,\label{photoncond}
\ee
where a prime denotes a derivative with respect to $r$. The first two conditions yield
\bea
\fft{L}{E} &\equiv& \xi=\frac{X\Delta'-2\Delta X'}{a\Delta'}\Big|_{r=r_{\rm ph}},\nn\\
\fft{Q}{E^2} &\equiv&\eta=\frac{4a^2X'^2\Delta-((X-a^2)\Delta'-2X'\Delta)^2}{a^2\Delta'^2}\Big|_{r=r_{\rm ph}}.\label{xieta1}
\eea
One constraint for the value of the photon¡¯s circular orbit radius is $\Theta\ge0$; furthermore, the requirement that $Q\ge 0$ restricts the range of $r_{\rm ph}$. In other words, for some specific choice of integration constant ratios $(\xi, \eta)$, the null geodesics can form close orbits, giving rise to photon surfaces.  In Einstein gravity satisfying at least null energy condition, the photon orbits are typically unstable.  Black holes with stable photon orbits are extremely rare \cite{Cvetic:2016bxi}. The only example of black holes that admit stable photon orbits was recently constructed satisfying the dominant energy condition \cite{Liu:2019rib}.  All the examples we present in this paper have no stable photon orbit outside the event horizon.

\subsubsection{Class-two metric ansatz}

In this paper, we also propose an alternative description of rotating black holes. The metric ansatz takes the new form
\be
ds^2 = -\fft{\rho^2}{\sqrt{W}}(dt + {\cal A} d\phi)^2 + \sqrt{W}
\left(\fft{dr^2}{\Delta} + d\theta^2 + \fft{\Delta \sin^2\theta}{\rho^2} d\phi^2\right),\label{classtwo}
\ee
where
\bea
\Delta &=& r^2 + a^2 - 2m r\,,\qquad \rho^2 = r^2 + a^2 \cos^2\theta -2m r\,,\nn\\
{\cal A} &=&  \fft{a \sin^2\theta}{\rho^2}{\cal B}(r)\,,\qquad W = {\cal B}(r)^2 + \rho^2 (U(r) + a^2 \cos^2\theta)\,,
\eea
It is important to emphasize that $\Delta$, ${\cal B}$ and $U$ are functions of the radial variable $r$ only. The ansatz can tolerate a constant shift of the radial coordinate, where $\Delta$ and $\rho^2$ acquire a constant.  In presenting the above, we have chosen the radial coordinate such that $\Delta$ and $\rho^2$ are bare minimum. The metric is then completely specified by the ${\cal B}$ and $U$ functions.  It is worth pointing out that the horizon geometry can be abstractly analysed. The horizon $r_+$ is the larger root of $\Delta$ and the entropy is given by
\be
S=\pi {\cal B}(r_+)\,.\label{entropy2}
\ee
leading to the mean radius of the horizon
\be
R_+=\sqrt{{\cal B}(r_+)}\,.
\ee
Note that the inequality (\ref{sGthetatheta}) is also satisfied by our class-two metrics.

We now compare the two metric Ans\"atze (\ref{classone}) and (\ref{classtwo}).  In the special case with
\be
U=\Delta + 2 {\cal B} -a^2\,,\label{UcalB}
\ee
the metric can be reduced to the previous one, with
\be
H=r^2 - 2m r + {\cal B} + a^2 \cos^2\theta\,,\qquad F=G=\fft{\rho^2}{H}\,,\qquad
X=\Delta + {\cal B}\,.
\ee
Note that the two entropy formulae (\ref{entropy1}) and (\ref{entropy2}) yield the same result in this case since $\Delta=0$ on the horizon. Conversely, when $F=G$ and the constraints in (\ref{DeltaX}) are satisfied, then the metric (\ref{classone}) can be put in the form of (\ref{classtwo}).  However, if $F\ne G$, then the metric (\ref{classone}) cannot be put into the form of (\ref{classtwo}).  Likewise, if the functions $U$ and ${\cal B}$ are not related by (\ref{UcalB}), then the metric (\ref{classtwo}) cannot be put into the form of (\ref{classtwo}).  In this paper, all the explicit examples we consider can be put into the form of (\ref{classtwo}), but not necessarily of the form of (\ref{classone}).

The equations of the null geodesic motions are
\bea
\sqrt{W} \fft{dt}{d\tau} &=& E (U + a^2 \cos^2\theta) + \fft{\cal B}{\Delta} (E {\cal B} - a L)\,,\nn\\
\sqrt{W} \fft{d\phi}{d\tau} &=& \fft{a (E B- aL)}{\Delta} + \fft{L}{\sin^2\theta}\,,\nn\\
\sqrt{W} \fft{dr}{d\tau} &=&\pm \sqrt{{\cal R}(r)}\,,\qquad {\cal R}(r)=(E{\cal B} - aL)^2 -\Delta\, (Q + L^2 - E^2 U),\nn\\
\sqrt{W} \fft{d\theta}{d\tau} &=&\pm\sqrt{\Theta(\theta)}\,,\qquad
\Theta(\theta) = Q+ Ea^2 \cos^2\theta + L^2\cot^2\theta\,.\label{geodesic2}
\eea
The photon orbits are again determined by (\ref{photoncond}) and therefore we have
\bea
\xi &=&\fft{{\cal B} \Delta' - \Delta ({\cal B}' + \sqrt{{\cal B}'^2 + \Delta' U'})}{a\Delta'}\Big|_{r=r_{\rm ph}},\nn\\
\eta &=& \fft{\Delta U -2a {\cal B} \xi - (\Delta-a^2)\xi^2}{\Delta}\Big|_{r=r_{\rm ph}}.\label{xieta2}
\eea
The existence of a square root in the above expression makes it much more difficult to analyse the photon orbits and the black hole shadows.  For special case (\ref{UcalB}), the determinant becomes a total square, namely
\be
{\cal B}'^2 + \Delta' U'=\ft14 (U' + \Delta')^2.
\ee
However, in this case, as mentioned earlier, the metric can be cast into the form of (\ref{classone}).  Nevertheless, as we shall see section \ref{sec:CY} that black holes in the class-two form do exist and we have to cope with the equations (\ref{xieta2}) in order to study the photon orbits and shadows.

\subsection{Black hole shadows}

Having obtained the condition for unstable photon orbits, we are in the position to study the photon shadows surrounded by photons escaped from the unstable orbits. For the asymptotically flat spacetime, the observer's sky is the celestial plane perpendicular to the line joining the observer at infinity and the center of the black hole. The celestial coordinates $x$ and $y$ are defined by \cite{Vazquez:2003zm}
\bea
x&=&\lim_{r_0\rightarrow\infty}\left(-r_0^2\sin\theta_0\frac{d\phi}{dr}\Big|_{(r_0,\theta_0)}\right),\\
y&=&\lim_{r_0\rightarrow\infty}\left(r_0^2\frac{d\theta}{dr}\Big|_{(r_0,\theta_0)}\right),
\eea
where $(r_0,\theta_0)$ are the position coordinates of the observer. To be specific, $\theta_0$ is the angle between the line and the black hole angular momentum vector, with $\theta_0=0$ corresponding to the north pole direction. Making use of the geodesic equations (\ref{geodesic1}) or (\ref{geodesic2}), we have
\be
x=-\xi\csc\theta_0\,,\qquad
y=\pm\sqrt{\eta+a^2\cos^2\theta_0-\xi^2\cot^2\theta_0}\,.\label{xy}
\ee
In other words, the photon observed in the celestial coordinates is determined by the integration constant ratios $(\xi,\eta)$ of the null geodesic motion.  In particular, we have
\be
x^2 + y^2 = \fft{L^2 + Q}{E^2} + a^2 \cos^2\theta_0\,.
\ee
For the given integration constants $(E,L,Q)$, the photons form a circle in the sky's plane, and its radius depends also on the viewer's angle.

However, we are considering a collection of photons with all allowed $(E,L,Q)$. It is clear that the unstable circular photon orbits form the boundary of a shadow, since those inside the photon surfaces will be trapped and cannot escape to infinity. The integration constant ratios $(\xi,\eta)$ of the unstable photon orbits are determined by the radial variable $r_{\rm ph}$ by (\ref{xieta1}) for class-one metrics (\ref{classone}) and by (\ref{xieta2}) for class-two metrics (\ref{classtwo}).  Thus the boundary of the shadow in the observer's sky is determined by the parametric functions in terms of the parameter $r_{\rm ph}$, namely
\be
x=x(r_{\rm ph})\,,\qquad y=y(r_{\rm ph})\,.
\ee
The allowed range of $r_{\rm ph}$ is restricted by requiring that $Q\ge 0$. We can then determine the shape of the shadow by eliminating the parameter $r_{\rm ph}$ and obtain the function of the closed loop $f(x,y)=0$, in the sky's plane.

The shape of the shadow depends on the observer's viewing angle $\theta_0$. When viewed from north pole $\theta =0$ (or equivalent south pole $\theta=\pi$,) the shadow remains a round disk. The photons that form the shadow boundary, a round circle, satisfy
\be
\xi(r^0_{\rm ph})=0\,,
\ee
and the shadow radius is
\be
R_{\rm sh}=\sqrt{a^2 + \eta(r^0_{\rm ph})}\,.\label{Rsh1}
\ee

When $\theta\ne 0$ or $\pi$, the shadows are no longer round, but distorted.  For given angle $\theta_0$, the maximum distortion occurs when the black hole has the maximum allowed angular momentum. For given mass and angular momentum, the distortion is largest when viewed from the $\theta_0=\ft12\pi$ equatorial plane. Roughly speaking, the vertical $y$-direction is elongated while the horizontal $x$-direction is squeezed, but the shadows remain convex. For $y=0$, there are two real solutions
\be
\eta(r_{\rm ph}^\pm)=0\,,\qquad \hbox{with}\qquad r_{\rm ph}^+\ge  r_{\rm ph}^-
\ee
We define the size of the shadow by
\be
R_{\rm sh} = \ft12 \big(x(r_{\rm ph}^+) - x(r_{\rm ph}^-)\big)\,.\label{Rsh2}
\ee
It is worth noting that typically the photons at $x^+$ and $x^-$ come from the orbits in the opposite or the same rotating direction of the black hole respectively. To understand this $R_{\rm sh}$ definition, it is instructive to examine the shadow plots presented in appendix \ref{sec:app-shadow}. We note that the boundary of the shadow is a closed convex loop. For any given point on the loop, we can find another point in the loop so that the line joining them is longest (in the Euclidean sense.)  To be precise, for any given point, there is a diagonal line with {\it local extremum} length. It can be argued based on the the symmetry that the $R_{\rm sh}$ defined above is the shortest of all these diagonal lines. For this reason, we choose $R_{\rm sh}$ to characterize the size of the photon shadow.

For general $\theta_0$, we determine $r_{\rm ph}^\pm$ by requiring
\be
y(r, \theta_0)\Big|_{r=r_{\rm ph}^\pm}=0\,.
\ee
where $y$ is given by (\ref{xy}).  It is important that the roots of the above equation $r_{\rm ph}^\pm$ must be chosen that both are outside of the horizon. When $r_{\rm ph}^-$ is inside of the horizon, then we need to define $r_{\rm ph}^-=r_+$.  When $r_{\rm ph}^\pm$ coincide, the shadow becomes round sphere.  The shadow size is then again formally given by (\ref{Rsh2}). When $\theta_0=0$, the quantity (\ref{Rsh2}) reduces to (\ref{Rsh1}).  All the results reduce to the same radius of the round disk, independent of the view angle $\theta_0$, when the black hole is spherically symmetric.  The photon shadows of the extremal Kerr black hole, viewed from $\theta_0=0$, $\theta_0=\pi/4$ and $\theta_0=\pi/2$, are presented in Fig.~\ref{fig1} in appendix \ref{sec:app-shadow}.  While we studied a large number of black holes, we present only the shadow plots for the Kerr metrics since the shadow shapes are all similar.

The characterization of the size of the whole photon surfaces is much less obvious, since they form regions
instead of just a thin shell \cite{Grenzebach:2014fha}. We note that the shadows observed at infinity are associated with specific photon orbits and hence it is only natural to consider the the size of these relevant photon orbits.  In other words, the photon orbits associated with the
photons around the edge of the shadow also depends on the observer's angle $\theta_0$.  In particular, the photons seen at $(x,y)=(x^\pm, 0)$ in the sky's plane are related to the photon orbits $r_{\rm ph}^\pm$. Therefore, we propose
\be
R_{\rm ph}=\fft12\Big(\sqrt{g_{\theta\theta}(r_{\rm ph}^+,\theta_0)}+\sqrt{g_{\theta\theta}(r_{\rm ph}^-,\theta_0})\Big),
\ee
to characterize the size of this specific photon orbit.

As was already calculated earlier, the effective mean radius of the black hole horizon on the other hand is independent of the location of the observation, and it is given by
\be
R_+=\sqrt{\frac{S}{\pi}}=
\left\{
  \begin{array}{ll}
    \sqrt{X(r_+)}, & \hbox{class one with $F=G$,} \\
    \sqrt{\cal B(r_+)}, & \hbox{class two.}
  \end{array}
\right.
\ee
We choose this to measure the size of the horizon also because it is relevant to the Penrose's black hole entropy bound. One may argue that the $\theta_0$-dependent radius, $\sqrt{g_{\theta\theta}(r_+,\theta_0)}$, should be the relevant horizon size.  It follows from the inequality (\ref{sGthetatheta}) that we have
\be
R_+\ge \sqrt{g_{\theta\theta}(r_+,\theta_0)}\,.
\ee
Our main conclusion of (\ref{conjecture}) will thus remain even if we choose $\sqrt{g_{\theta\theta}(r_+,\theta_0)}$. In this paper we opt to choose the entropy related mean radius as our $R_+$, to arrive at the stronger inequality bound.

When the black hole is spherically symmetric, the black hole horizon, photon sphere and shadow disk are all round, and the above quantity $R_+$, $R_{\rm ph}$ and $R_{\rm sh}$ become the standard radii of these respective geometric shapes.  A sequence of inequalities (\ref{conjecture}) was proposed and validated with large number of explicit examples. In this paper, with our generalizations of those variables to incorporate stationary rotating black holes, we conjecture that they continue to be valid. In order to verify the inequalities, we define
\be
{\cal X}=\fft{3\sqrt3M}{R_{\rm sh}}\,,\qquad {\cal Y}=\fft{R_{\rm sh}}{\sqrt3 R_{\rm ph}}\,,\qquad
{\cal Z}=\fft{2 R_{\rm ph}}{3R_+}\,,\label{XYZ}
\ee
and verify that ${\cal X}\ge 1$, ${\cal Y}\ge 1$ and ${\cal Z}\ge 1$.  In other words, for given mass $M$, the bigger the $({\cal X,Y,Z})$, the smaller is the black hole. The general proof appears to be formidable, and we shall verify them with explicit examples in the subsequent sections.  We shall focus on two viewing angles: $\theta_0=0$, for which the shadow is a round disk, and $\theta_0=\ft12\pi$, for which the shadow is maximally distorted from a round disk.

\section{Kerr black hole}

We first consider the simplest case, namely the Kerr black hole \cite{Kerr:1963ud}.  The metric can be put into (\ref{classone}) with
\be
F=G=1-\frac{2mr}{\rho^2},\qquad H=r^2+a^2\cos^2\theta=\rho^2\,.
\ee
Indeed, the functions $X$ and $\Delta$ depend on $r$ only, given by
\be
X=r^2+a^2,\qquad \Delta=r^2-2mr+a^2.
\ee
It can also be put into the form of (\ref{classtwo}), with
\be
U=r^2 + 2m r\,,\qquad {\cal B}=2m r\,.
\ee
The solution has mass $M=m$ and when $m\ge a$, the metric describes a black hole and the horizons are determined by $\Delta=0$, which admits two roots, corresponding to the inner and outer horizons.
The event horizon, i.e. the outer horizon, is located at $r_+ =m + \sqrt{m^2-a^2}$.  The black hole entropy and the mean horizon radius are
\be
S=\pi X(r_+)=\pi (r_+^2 + a^2)\,,\qquad R_+=\sqrt{r_+^2 + a^2}\,.
\ee

The geodesic motions around the Kerr black hole were first discussed in \cite{Bardeen:1973tla}.
Following from the general discussion in section \ref{sec:photon}, we have
\bea
\xi&=&\frac{(3m-r_{\rm ph})r_{\rm ph}^2-a^2(m+r_{\rm ph})}{a(r_{\rm ph}-m)},\\
\eta&=&\frac{r_{\rm ph}^3(4a^2m-r_{\rm ph}(3m-r_{\rm ph})^2)}{a^2(r_{\rm ph}-m)^2}.
\eea
There is clearly an upper bound of $r_{\rm ph}$ since we must have $\eta\ge 0$. The shape of the shadow depends on the angle $\theta_0$ of the observer.  We shall focus on two cases, namely $\theta_0=0$, corresponding to the viewing from the north pole, or $\theta=\pi/2$, corresponding to the viewing from the equatorial plane.  We also present some discussions on the shadows of general $\theta_0$.

\subsection{$\theta_0=0$}

In this case, an observer sees a round disk shadow surrounded by photons originated from the round unstable photon orbits. The radius of the photon orbits is determined by $\xi(r_{\rm ph})=0$, namely
\be
a^2 (m+r_{\rm ph})+r_{\rm ph}^2 (r_{\rm ph}-3 m)=0\,.
\ee
It admits in general three real solutions for applicable $(m,a)$. The root outside the horizon is the largest and it is given by
\be
r_{\rm ph} =m+2 \sqrt{m^2-\frac{a^2}{3}} \cos \left(\frac{1}{3} \cos ^{-1}\left(\frac{3 \sqrt{3} m \left(m^2-a^2\right)}{\left(3 m^2-a^2\right)^{3/2}}\right)\right).
\ee
Following from the discussion in section \ref{sec:photon}, we have
\be
R_{\rm sh} = \sqrt{\fft{2m(3r_{\rm ph}^2 - a^2)}{r_{\rm ph}-m}}\,,\qquad R_{\rm ph} = \sqrt{r_{\rm ph}^2 + a^2}\,.
\ee
The functions $({\cal X},{\cal Y},{\cal Z})$ defined by (\ref{XYZ}) in this case depend only on the dimensionless quantity $\lambda=a/m$, given by
\bea
{\cal X}^2&=& \frac{27 \sqrt{1-\frac{\lambda ^2}{3}} \cos \left(\frac{1}{3} \cos ^{-1}\left(\frac{3 \sqrt{3} \left(1-\lambda ^2\right)}{\left(3-\lambda ^2\right)^{3/2}}\right)\right)}{\frac{1}{3} \left(2 \sqrt{9-3 \lambda ^2} \cos \left(\frac{1}{3} \cos ^{-1}\left(\frac{3 \sqrt{3} \left(1-\lambda ^2\right)}{\left(3-\lambda ^2\right)^{3/2}}\right)\right)+3\right)^2-\lambda ^2}\,,\nn\\
{\cal Y}^2&=& \frac{\left(\frac{1}{3} \left(2 \sqrt{9-3 \lambda ^2} \cos \left(\frac{1}{3} \cos ^{-1}\left(\frac{3 \sqrt{3} \left(1-\lambda ^2\right)}{\left(3-\lambda ^2\right)^{3/2}}\right)\right)+3\right)^2-\lambda ^2\right)}{3 \sqrt{1-\frac{\lambda ^2}{3}} \left(\lambda ^2+\left(2 \sqrt{1-\frac{\lambda ^2}{3}} \cos \left(\frac{1}{3} \cos ^{-1}\left(\frac{3 \sqrt{3} \left(1-\lambda ^2\right)}{\left(3-\lambda ^2\right)^{3/2}}\right)\right)+1\right)^2\right)}\nn\\
&&\times \sec \left(\frac{1}{3} \cos ^{-1}\left(\frac{3 \sqrt{3} \left(1-\lambda ^2\right)}{\left(3-\lambda ^2\right)^{3/2}}\right)\right),\nn\\
{\cal Z}^2 &=& \frac{2 \left(\lambda ^2+\left(2 \sqrt{1-\frac{\lambda ^2}{3}} \cos \left(\frac{1}{3} \cos ^{-1}\left(\frac{3 \sqrt{3} \left(1-\lambda ^2\right)}{\left(3-\lambda ^2\right)^{3/2}}\right)\right)+1\right)^2\right)}{9 \left(\sqrt{1-\lambda ^2}+1\right)}\,.
\eea
Here the dimensionless constant $\lambda$ lies in region $[0,1]$, with $\lambda=0, 1$ corresponding to the Schwarzschild black hole and the extremal rotating black hole respectively.
We find that $({\cal X}, {\cal Y}, {\cal Z})$ are all monotonically increasing function as the parameter $\lambda$ runs from 0 to 1. We do not have a clever analytical way to demonstrate this, but it can be easily seen by numerical plots. For small $\lambda$, we have
\be
\{{\cal X},{\cal Y},{\cal Z}\} = \{1 + \ft1{18} \lambda^2, 1 +\ft1{27} \lambda^2, 1 + \ft7{216} \lambda^2\} + {\cal O}(\lambda^4)\,.
\ee
Near $\lambda=1$, we have
\bea
{\cal X} &=&\frac{3\sqrt{3}}{2} \left(\sqrt{2}-1\right)  +\frac{3 \sqrt{3}}{4} \left(2 \sqrt{2}-3\right) (1-\lambda)
+{\cal O}\left((1-\lambda)^2\right)\,,\nn\\
{\cal Y}&=& \sqrt{\ft{1}{3} \left(2+\sqrt{2}\right)}-\frac{1}{4} \sqrt{\ft{1}{3} \left(2+\sqrt{2}\right)} (1-\lambda) +O\left((1-\lambda)^2\right)\,,\nn\\
{\cal Z}&=& \frac23 {\sqrt{2+\sqrt{2}}}-\frac{1}{3} \sqrt{2 \left(2+\sqrt{2}\right)} \sqrt{1-\lambda}+O\left(1-\lambda\right)\,.
\eea
Thus we have ${\cal X}\ge1$, ${\cal Y}\ge1$ and ${\cal Z}\ge 1$.

Since all $({\cal X},{\cal Y},{\cal Z})$ functions are monotonically increasing with respect to $\lambda$, it follows that for given mass, the Schwarzschild is the biggest and the bigger the angular momentum, all the size parameters $(R_+, R_{\rm ph}, R_{\rm sh})$ becomes smaller. When the black hole becomes extremal, all size parameters become the minima, with ${\cal X}=1.0762$, ${\cal Y}=1.0668$ and ${\cal Z}=1.2318$.  We also see that while the distortion of the horizon is significant in the extremal limit, the size of the photon shadow is largely unchanged.

\subsection{$\theta_0=\pi/2$}

The shadow is skewed most when viewed from the equatorial plane at $\theta_0=\fft12 \pi$.  For example, in the extremal limit, the shape of the shadow resembles a filled letter D. The circular orbit radii of the photons on the equatorial plane is given by $\eta=0$
\be
r_{\rm ph}^{\pm}=2m\left[1+\cos\left(\frac{2}{3}\arccos\Big(\pm\frac{a}{m}\Big)\right)\right]
\ee
\be
x^\pm = -\xi|_{r=r_{\rm ph}^\pm}\,.
\ee
We follow the discussion in section \ref{sec:photon} and define
\be
R_{\rm ph} = \ft12 (r_{\rm ph}^+ + r_{\rm ph}^-)\,,\qquad R_{\rm sh} = \ft12(x^+ - x^-)\,.
\ee
We can now again evaluate the functions $({\cal X},{\cal Y},{\cal Z})$ defined by (\ref{XYZ}).  They are again functions of the dimensionless parameter $\lambda$ and the explicit expressions are
\bea
{\cal X} &=& \frac{3 \sqrt{3} \lambda \left(2 C_-+1\right) \left(2 C_++1\right) }{2 \left(C_+-C_-\right) \left(2 C_+^2+3 C_++C_-^2 \left(4 C_++2\right)+C_- \left(4 C_+^2+8 C_++3\right)-\lambda ^2+2\right)}\,,\nn\\
{\cal Y}&=&\frac{2 \left(C_+-C_-\right) \left(2 C_+^2+3 C_++C_-^2 \left(4 C_++2\right)+C_- \left(4 C_+^2+8 C_++3\right)-\lambda ^2+2\right)}{\sqrt{3}\lambda \left(2 C_-+1\right) \left(2 C_++1\right)\left(C_-+C_++2\right) }\,,\nn\\
{\cal Z}&=&\frac{\sqrt{2} \left(C_-+C_++2\right)}{3 \sqrt{\sqrt{1-\lambda ^2}+1}}\,,
\eea
where
\be
C_\pm = \cos \left(\frac{2}{3} \cos ^{-1}(\pm\lambda )\right)\,.
\ee
It is easy to verify using numerical plots that $\cal (X,Y,Z)$ are also all monotonically increasing function of the dimensionless parameter $\lambda\in [0,1]$.  In other words, for given mass, the larger the angular momentum, the smaller are the size parameters. Near $\lambda=0$, we have
\be
\left\{{\cal X},{\cal Y},{\cal Z}\right\}=\left\{1+\frac{\lambda^2}{18},1+\frac{\lambda^2}{54},1+\frac{11 \lambda^2}{216}\right\} + {\cal O}(\lambda^4)\,.
\ee
Near $\lambda=1$, we have
\bea
\left\{{\cal X},{\cal Y},{\cal Z}\right\}&=&\left\{\frac{2}{\sqrt{3}}-\frac{2}{9} \sqrt{2} \sqrt{1-\lambda},\frac{3 \sqrt{3}}{5}-\frac{1}{25} \sqrt{2} \sqrt{1-\lambda},\frac{5}{3 \sqrt{2}}+\left(\frac{2}{3 \sqrt{3}}-\frac{5}{6}\right) \sqrt{1-\lambda}\right\}\nn\\
&& + {\cal O}(1-\lambda)\,.
\eea
Thus we have ${\cal X}\ge 1$, ${\cal Y}\ge 1$ and ${\cal Z}\ge 1$ again.

\subsection{General $\theta_0$}

For general angle $\theta_0$, the equation $y=0$ that determines $r_{\rm ph}^\pm$ becomes a polynomial of six order in terms of the radial coordinate $r$ and hence an analytic solution for $r_{\rm ph}^\pm$ are no longer possible.  For small $a$, we can obtain $r_{\rm ph}^\pm$ as the Taylor expansion:
\be
\fft{r_{\rm ph}^\pm}{m} = 3 \pm \fft2{\sqrt3} \sin\theta_0\, \lambda -\fft19 (3 + \cos2\theta_0)
\lambda^2 \pm \fft{13\sin\theta_0 + 3\sin3\theta_0}{54\sqrt3} \lambda^3 + {\cal O}(\lambda^4)\,,
\ee
with $\lambda=a/m$. This allows us to calculate $R_{\rm ph}$ and $R_{\rm sh}$ and eventually $({\cal X},{\cal Y}, {\cal Z})$ defined in (\ref{XYZ}). We find that for small $a$
\be
\{{\cal X, Y,Z}\} = \left\{1 + \ft1{18} \lambda^2, 1 + \ft{1}{108} (3 + \cos2\theta_0)\lambda^2, 1 + \ft{1}{216} ( 9 - 2 \cos2\theta_0) \lambda^2\right\} + {\cal O}(\lambda^3)\,.
\ee
When $\theta_0=0$ and $\theta_0=\pi/2$, these quantities reduce to those in the previous subsections. Note that we have to obtain $r_{\rm ph}^\pm$ up to and including the cubic order of $\lambda$ in order to obtain the expressions for ${\cal X,Y}$ and $\cal Z$ at the quadratic order.  In the extremal limit with $\lambda=1$, the $y(r_{\rm ph})=0$ equation is an quartic polynomial of $r_{\rm ph}$ and hence it can be solved analytically.  A numerical plot indicates that all $\cal (X,Y,Z)$ are again bigger than 1 in this case.  The subtlety in the extremal limit is that when $\theta_0$ approaches the equatorial plane, $r_{\rm ph}^-$ solved from $y=0$ can be smaller than the horizon radius and hence it should be replaced by $r_+$.  For general $\theta_0$ and $\lambda$, we verified the inequalities for an incomplete but large number of numerical data.

\section{Kerr-Newman black hole}

The Kerr-Newman black hole \cite{Newman:1965tw,Newman:1965my} can be cast in the forms of both the class-one (\ref{classone}) and class-two (\ref{classtwo}) metrics.  In this section, we shall use the class-one metric. The metric functions are given by
\be
F=G=1-\frac{2mr}{\rho^2}+\frac{q^2}{\rho^2},\qquad H=r^2+a^2\cos^2\theta=\rho^2\,.
\ee
Indeed, both $X$ and $\Delta$ are functions of $r$ only, given by
\be
X=r^2+a^2,\qquad \Delta=r^2-2mr+a^2+q^2.
\ee
The horizon mean radius is
\be
R_{+}=\sqrt{r_+^2 + a^2}\,,\qquad r_+=m+ \sqrt{m^2-a^2-q^2}\,.
\ee
It follows from the discussion in section \ref{sec:photon} that the unstable photon orbits satisfy
\bea
\xi&=&\frac{2r_{\rm ph}(2mr_{\rm ph}-q^2)-(r_{\rm ph}+m)(r_{\rm ph}^2+a^2)}{a(r_{\rm ph}-m)}\,,\nn\\
\eta&=&\frac{4a^2r_{\rm ph}^2(mr_{\rm ph}-q^2)-r_{\rm ph}^2(r_{\rm ph}(r_{\rm ph}-3m)+2q^2)^2}{a^2(r_{\rm ph}-m)^2}\,.
\eea
We can now determine the photon shadows.  We shall focus only on the $\theta_0=0$ and $\theta_0=\pi/2$ cases.

\subsection{$\theta_0=0$}

The radius of the relevant photon orbits $R_{\rm ph}=\sqrt{r_{\rm ph}^2 + a^2}$ is determined by the largest root of the cubic polynomial
\be
r_{\rm ph}^3-3m r_{\rm ph}^2 + (a^2 + 2q^2) r_{\rm ph} + a^2 m =0\,.
\ee
The largest root is
\be
r_{\rm ph}=m + 2 \sqrt{m^2-\ft13 (a^2+2q^2)} \cos \left(\frac{1}{3} \cos ^{-1}\left(\frac{3 \sqrt{3} m \left(m^2-a^2 - q^2\right)}{\left(3 m^2 -a^2-2 q^2\right)^{3/2}}\right)\right).
\ee
The shadow remains a round disk, and the radius is
\be
R_{\rm sh} = \sqrt{\fft{6m r_{\rm ph}^2 - 4 q^2 r_{\rm ph} - 2 a^2 m}{r_{\rm ph} - m}}\,.
\ee
We are now in the position to write the $\cal (X,Y,Z)$ functions.  We introduce the dimensionless parameters
$(\lambda, \sigma)$ by
\be
\fft{a}{m}=\lambda \sqrt{1-\sigma^2}\,,\qquad \fft{q}{m} = \lambda\sigma\,.
\ee
Both parameters lie in the region $[0,1]$.  We find
\bea
{\cal X}^2 &=& \frac{27}{\sqrt{9-3\lambda ^2 \left(\sigma ^2+1\right)}\,(4C + C^{-1})-4 \lambda ^2 \sigma ^2+12}\,,\nn\\
{\cal Y}^2 &=& \frac{\sqrt{9-3\lambda ^2 \left(\sigma ^2+1\right)}(4C + C^{-1})-4 \lambda ^2 \sigma ^2+12}{4 C^2 \left(3-\lambda ^2 \left(\sigma ^2+1\right)\right)+4C \sqrt{9-3\lambda ^2 \left(\sigma ^2+1\right)}+3
\lambda ^2 \left(1-\sigma ^2\right)+3}\,,\nn\\
{\cal Z}^2 &=& \frac{4 \left(\left(\frac23 C \sqrt{9-3\lambda ^2 \left(\sigma ^2+1\right)}+1\right)^2+\lambda ^2 \left(1-\sigma ^2\right)\right)}{9 \left(2-\lambda ^2 \sigma ^2+2 \sqrt{1-\lambda ^2}\right)}\,.
\eea
where
\be
C\equiv \cos \left(\frac{1}{3} \cos ^{-1}\left(\frac{3 \sqrt{3} \left(1-\lambda ^2\right)}{\left(3-\lambda ^2 \left(\sigma ^2+1\right)\right)^{3/2}}\right)\right).
\ee
The expression becomes much simpler in the extremal $\lambda=1$ limit, in which case, we have
\be
{\cal X}^2= \frac{27}{12-4 \sigma ^2+8 \sqrt{2-\sigma ^2}},\quad
{\cal Y}^2= \frac{2}{3} \left(1+\frac{1}{\sqrt{2-\sigma ^2}}\right),\quad
{\cal Z}^2=\frac{8}{9} \left(1+\frac{1}{\sqrt{2-\sigma ^2}}\right).
\ee
Note that $\sigma=0$ leads to the Kerr result.  $\sigma=1$ leads to the RN black hole.  Our numerical contour plots indicate that ${\cal X}\ge 1$, ${\cal Y}\ge 1$ and ${\cal Z}\ge 1$ for all $\lambda, \sigma \in [0,1]$.  We also notice that for fixed mass $m=1$ and charge $q=\lambda \sigma$, the bigger than angular momentum, the smaller is the black hole size.

\subsection{$\theta=\fft12\pi$}

In this case, we have to deal with a quartic equation to obtain $r_{\rm ph}$
\be
r_{\rm ph}^4-6mr^3+(9m^2+4q^2)r_{\rm ph}^2-4m(a^2+3q^2)r_{\rm ph}+4q^2(a^2+q^2)=0\,.
\ee
The equation can be solved exactly and the subtlety is to select the right roots that are outside the horizon.  The general formula is presented in appendix \ref{sec:quartic}.  The quantities $\cal (X,Y,Z)$ are again functions of dimensionless parameters $(\lambda, \sigma)$ only.  The expressions are too messy to present; however, we can perform exhaustive numerical plots for the parameters $(\lambda, \sigma)\in [0,1]$.  Our contour plots indicate that quantities $\cal (X,Y,Z)$ are again no smaller than 1.

In the extremal limit $\lambda=1$, the situation is much simpler and we have
\bea
\fft{r_{\rm ph}^+}{m} &=& 2\left(1 + \sqrt{1-\sigma^2}\right)\,,\nn\\
\fft{r_{\rm ph}^-}{m} &=&
\left\{
  \begin{array}{ll}
    1, &\qquad 0\le \sigma\le \fft{\sqrt3}{2},\\
    2\left(1 - \sqrt{1-\sigma^2}\right), &\qquad \fft{\sqrt3}{2}\le\sigma<1\,.
  \end{array}
\right.
\eea
We can then obtain the shadow size parameter:
\be
\fft{R_{\rm sh}}{m} =
\left\{
  \begin{array}{ll}
   2+ \frac{5-4 \sigma ^2}{2 \sqrt{1-\sigma ^2}}, &\qquad  0\le \sigma\le \fft{\sqrt3}{2}\, \\
    4, & \qquad \fft{\sqrt3}{2}<\sigma\le 1\,.
  \end{array}
\right.
\ee
We therefore have explicit analytical expressions for $\cal (X,Y,Z)$, namely
\bea
{\cal X} &=&
\left\{
  \begin{array}{ll}
   \frac{6 \sqrt{3-3 \sigma ^2}}{-4 \sigma ^2+4 \sqrt{1-\sigma ^2}+5} , &\qquad  0\le \sigma\le \fft{\sqrt3}{2}\, \\
  \frac{3 \sqrt{3}}{4}, & \qquad \fft{\sqrt3}{2}<\sigma\le 1\,.
  \end{array}
\right.,\nn\\
{\cal Y} &=&
\left\{
  \begin{array}{ll}
  \frac{\sqrt{3} \left(-4 \sigma ^2+4 \sqrt{1-\sigma ^2}+5\right)}{-6 \sigma ^2+9 \sqrt{1-\sigma ^2}+6}, &\qquad  0\le \sigma\le \fft{\sqrt3}{2}\, \\
   \frac{2}{\sqrt{3}}, & \qquad \fft{\sqrt3}{2}<\sigma\le 1\,.
  \end{array}
\right.\nn\\
{\cal Z} &=&
\left\{
  \begin{array}{ll}
   \frac{2 \sqrt{1-\sigma ^2}+3}{3 \sqrt{2-\sigma ^2}}, &\qquad  0\le \sigma\le \fft{\sqrt3}{2}\, \\
   \frac{4}{3 \sqrt{2-\sigma ^2}}, & \qquad \fft{\sqrt3}{2}<\sigma\le 1\,.
  \end{array}
\right.
\eea
These are all manifestly no smaller than 1 for $\sigma=[0,1]$.  In terms of the variable $\sigma$, the above functions are continuous at $\sigma=\sqrt3/2$, but not their derivatives with respect to $\sigma$.

\section{Kerr-Sen black hole}

The Kerr-Sen black hole \cite{Sen:1992ua} can also be written in both (\ref{classone}) and (\ref{classtwo}) coordinates.  We shall use the coordinate system of (\ref{classone}) here and the metric functions are
\be
F=G=\frac{\rho^2-r_1r}{\rho^2+r_2r},\quad H=\rho^2 +r_2 r,\quad \rho^2=r^2+a^2\cos^2\theta\,.
\ee
Consequently we have $X(r)$ and $\Delta(r)$, given by
\be
X=r^2+r_2r+a^2,\qquad \Delta=r^2-r_1r+a^2,
\ee
where $r_1$ and $r_2$ are related to the mass $M$ and electric charge $Q_e$ by
\be
r_1+r_2=2M\,,\qquad r_2=\fft{Q_e^2}M\,.
\ee
The entropy is $S=\pi X(r_+)$, where $r_+$ is the location of the outer horizon. Thus the mean radius of the horizon is
\be
R_+=\sqrt{M \sqrt{2 \left(2 M-r_2\right) \left(\sqrt{\left(2 M-r_2\right){}^2-4 a^2}+2 M-r_2\right)-4 a^2}}\,.
\ee
The radial coordinate of the photon orbits are determined by
\bea
\xi&=&\frac{a^2(2M+2r_{\rm ph}+r_2)+r_{\rm ph}(2r_{\rm ph}^2+3r_2r_{\rm ph}+r_2^2-2M(3r_{\rm ph}+r_2))}{a(2M-2r_{\rm ph}-r_2)}\,,\nn\\
\eta&=&-\frac{r_{\rm ph}^2(-8a^2M(2r_{\rm ph}+r_2)+(2r_{\rm ph}^2+3r_2r_{\rm ph}+r_2^2-2M(3r_{\rm ph}+r_2))^2)}{a^2(2M-2r_{\rm ph}-r_2)^2}\,.
\eea
For simplicity, we shall study the shadows viewed only from two angles, namely $\theta_0=0$ and $\pi/2$.

\subsection{$\theta_0=0$}

In this case, the size parameter of the relevant photon orbits is
\be
R_{\rm ph} = \sqrt{r_{\rm ph}^2 + a^2 + r_2 r_{\rm ph}}\,,
\ee
where $r_{\rm ph}$ is the largest roots of the cubic equation
\be
r_{\rm ph}^3 - \ft32 r_{\rm ph}^2 (2 M - r_2)  + \ft12 r_{\rm ph} (2 a^2 - 2 M r_2 + r_2^2)+
\ft12 a^2 (2 M + r_2)=0\,.
\ee
The solution is
\bea
r_{\rm ph}&=& M-\frac{r_2}{2}+\frac{1}{\sqrt{3}}\sqrt{12 M^2-4 a^2-8 M r_2+r_2^2}\nn\\
 &&\times\cos \left(\frac{1}{3} \cos ^{-1}\left(\frac{6 \sqrt{3} M \left(\left(r_2-2 M\right){}^2-4 a^2\right)}{\left(12 M^2-4 a^2-8 M r_2+r_2^2\right){}^{3/2}}\right)\right).
\eea
The radius of the round photon shadow is
\be
R_{\rm sh} = \frac{\sqrt{M \left(-2 a^2 \left(M+r_2\right)+r_{\rm ph} \left(-4 a^2+2 M r_2-r_2^2\right)+2 r_{\rm ph}^2 \left(3 M-r_2\right)\right)}}{+2 r_{\rm ph}+r_2 -2 M}\,.
\ee
It is convenient to introduce two dimensionless parameters $(\lambda, \sigma)$, defined by
\be
a=M\lambda\sigma\,,\qquad r_2 = 2M (1-\sigma)\,,\qquad \lambda, \sigma \in [0,1]\,,
\ee
for which $\Delta = (r-M\sigma)^2-M^2\sigma^2 (1-\lambda^2)$. We find that the quantities $\cal (X,Y,Z)$ of (\ref{XYZ}) are functions of $(\lambda,\sigma)$ only and the analytical expressions are
\bea
{\cal X}^2&=&\frac{27 C^2 \left(\sigma \tilde \lambda^2+2\right)}{4 C^2 (2 \sigma +1) \left(\sigma  \tilde \lambda^2+2\right)+4 \sqrt{3} C \sqrt{\sigma } \left(\sigma  \tilde \lambda^2+2\right)^{3/2}+9 \sigma  \tilde \lambda^2}\,,\nn\\
{\cal Y}^2&=&\frac{4 C^2 (2 \sigma +1) \left(\sigma  \tilde \lambda^2+2\right)+4 \sqrt{3} C \sqrt{\sigma } \left(\sigma  \tilde \lambda^2+2\right)^{3/2}+9 \sigma  \tilde \lambda^2}{C^2 \left(\sigma  \tilde \lambda^2+2\right) \left(\left(8 C^2+6\right) \sigma +\left(4 C^2-3\right) \sigma ^2 \tilde \lambda^2+4 \sqrt{3} C \sqrt{\sigma } \sqrt{\sigma  \tilde \lambda^2+2}\right)}\,,\nn\\
{\cal Z}^2&=&\frac{2 \left(\left(8 C^2+6\right) \sigma +\left(4 C^2-3\right) \sigma ^2 \tilde \lambda^2+4 \sqrt{3} C \sqrt{\sigma } \sqrt{\sigma  \tilde \lambda^2+2}\right)}{27 \sqrt{\sigma } \sqrt{\sigma +\sigma  \tilde \lambda^2+2 \sigma  \tilde \lambda}}\,.
\eea
where $\tilde \lambda=\sqrt{1-\lambda^2}$ and
\be
C=\cos \left(\frac{1}{3} \cos ^{-1}\left(\frac{3 \sqrt{3} \tilde\lambda^2 \sigma ^2}{\left( \tilde\lambda^2 \sigma^2 +2\sigma\right)^{3/2}}\right)\right).
\ee
These analytical expressions allow us to perform contour plots for all the parameters $(\lambda, \sigma)\in [0,1]$ exhaustively and demonstrate numerically that $\cal (X,Y,Z)$ are no less than 1. In the extremal limit, $\lambda=1$, the ratios all become much simpler, given by
\be
\left\{{\cal X,Y,Z}\right\}=\left\{\frac{3 \sqrt{\frac{3}{2}}}{2 \sqrt{\sigma }+\sqrt{2}},\,\sqrt{\frac{\sqrt{2}}{3\sqrt{\sigma }}+\fft23},\,\frac{2}{3} \sqrt{\frac{\sqrt{2}}{\sqrt{\sigma }}+2}\right\}>1\,.
\ee

\subsection{$\theta_0=\fft12\pi$}

The case of $\theta_0=\pi/2$ is more complicated, but simpler than the generic $\theta_0$.  The relevant $r_{\rm ph}^\pm$ satisfy the quartic equation
\bea
4r_{\rm ph}^4+12(r_2-2M)r_{\rm ph}^3+(36M^2-44Mr_2+13r_2^2)r_{\rm ph}^2\nn\\
-(16a^2M-6r_0(r_2-2M)^2)r_{\rm ph}-(8a^2M-r_2(r_2-2M)^2)r_2=0\,.
\eea
Using the formula in appendix \ref{sec:quartic}, we can obtain the correct roots and then determine both $R_{\rm ph}$ and $R_{\rm sh}$, following the description in section \ref{sec:photon}.  The quantities $\cal (X,Y,Z)$ are again functions of the dimensionless parameters $(\lambda, \sigma)$. The formulae in this case are all messy and we shall not present them but simply report the conclusion. We can contour plot these quantities and verify that they are indeed no less than 1.

     The inequalities can be manifestly established in the extremal $\lambda=1$ limit, in which case, we have
\be
r_{\rm ph}^+=2M(\sigma + \sqrt{\sigma})\,,\qquad r_{\rm ph}^-=M \sigma=r_+\,.
\ee
We can then read off the $R_{\rm ph}$ and $R_{\rm sh}$ and hence we have
\be
{\cal X}= \frac{6 \sqrt{3}}{\left(\sqrt{\sigma }+2\right)^2}\,,\quad
{\cal Y}= \frac{\left(\sqrt{\sigma }+2\right)^2}{\sqrt[4]{9\sigma } \left(\sqrt{2-\sigma } \sqrt[4]{\sigma }+2 \sqrt{\sigma }+2\right)}\,,\quad
{\cal Z}=\frac{\sqrt{2-\sigma } \sqrt[4]{\sigma }+2 \sqrt{\sigma }+2}{3 \sqrt[4]{4\sigma }}\,.
\ee
Thus we see analytically that these quantities are greater than 1 for the extremal black holes.

\section{Kerr-Cveti\v c-Youm black hole}
\label{sec:CY}

The 4-charge rotating black holes in the STU supergravity model \cite{Duff:1995sm} was constructed by Cveti\v c and Youm in \cite{Cvetic:1996kv}.  The typos were later corrected in \cite{Chong:2004na}.  We shall adopt the notations of \cite{Chong:2004na} in this paper. It should be emphasized that rotating black holes discussed in the previous sections are all special case of this general solutions.  The general metric cannot be cast into the form of (\ref{classone}) analytically.  We present the solution instead in the class-two metric (\ref{classtwo}), with the metric functions
\bea
{\cal B} &=&  2m \big(r c_{1234} - (r-2m) s_{1234}\big)\,,\nn\\
U&=& r^2 + 2m r(1 + s_1^2 + s_2^2 + s_3^2 + s_4^2) + 8m^2 (c_{1234}-s_{1234})s_{1234}\nn\\
&& -4m^2(s_{123}^2 + s_{124}^2 + s_{134}^2 + s_{234}^2)\,,
\eea
with
\be
c_{i_1\cdots i_n} = \cosh\delta_{i_1}\cdots  \cosh\delta_{i_n}\,,\quad
s_{i_1\cdots i_n} = \sinh\delta_{i_1}\cdots  \sinh\delta_{i_n}\,.
\ee
The mass of the solution is
\be
M=m\Big(1 + \ft12 (s_1^2 + s_2^2 + s_3^2 + s_4^2)\Big).
\ee
The four charges of the STU model are parameterized by $m$ and four dimensionless parameters $\delta_i$.
The solution describes a black hole when $m\ge a$, with the entropy given by
\be
S_+=\pi (r_+^2 + a^2)\big(c_{1234} + \fft{a^2}{r_+^2} s_{1234}\big)\,,\qquad R_+=\sqrt{\fft{S}{\pi}},
\ee
where $r_+=m + \sqrt{m^2 - a^2}$.

\subsection{Pairwise equal charges}

We first consider the case with pairwise equal charges, by setting $\delta_3=\delta_1$ and $\delta_4=\delta_2$.  The metric can be put in both class-one and class-two Ans\"atze.  The photon orbits are determined by
\bea
\xi &=& \fft{1}{a(m-r)}\Big(r^3 -3 m r^2 + r \left(a^2-2 m^2 \left(\left(2 s_2^2+1\right) s_1^2+s_2^2\right)\right)\nn\\
&&+a^2 m \left(2 s_1^2+2 s_2^2+1\right)+4 m^3 s_1^2 s_2^2\Big)\Big|_{r=r_{\rm ph}},\nn\\
\eta &=& \fft{1}{a^2(r-m)^2}\Big[-r^6 + 6m r^5 +m^2 \left(\left(8 s_2^2+4\right) s_1^2+4 s_2^2-9\right)r^4\nn\\
&&+ \left(4 a^2 m \left(s_1^2+s_2^2+1\right)-4 m^3 \left(\left(8 s_2^2+3\right) s_1^2+3 s_2^2\right)\right)r^3\nn\\
&&+4m^2\Big(a^2 \left(s_1^4+\left(6 s_2^2+1\right) s_1^2+s_2^4+s_2^2\right)\nn\\
&&\quad\qquad -m^2 \left(\left(2 s_2^2+1\right)^2 s_1^4+4 s_2^2 \left(s_2^2-1\right) s_1^2+s_2^4\right)\Big)r^2\nn\\
&&+\left(16 a^2 m^3 s_1^2 s_2^2 \left(s_1^2+s_2^2-1\right)+16 m^5 s_1^2 s_2^2 \left(\left(2 s_2^2+1\right) s_1^2+s_2^2\right)\right)r\nn\\
&&-16 a^2 m^4 s_1^2 s_2^2 \left(s_1^2+s_2^2\right)-16 m^6 s_1^4 s_2^4
\Big]\Big|_{r=r_{\rm ph}}.
\eea
Note that the numerator of $\eta$ is now a sextic polynomial of $r$, which makes it difficult to study the
situation analytically.

However, for $\theta_0=0$, the shadow remains a round disk and the relevant photon orbits are determined by
$\xi(r_{\rm ph})=0$, which is a cubic equation.  We thus have
\bea
\fft{r_{\rm ph}}{m} &=& 1+ {2 \gamma C} \,,\quad
\gamma = \sqrt{\ft13(3-\lambda ^2+2 s_1^2+2 s_2^2+4 s_1^2 s_2^2)}\,,\quad \lambda=\fft{a}{m}\,,\nn\\
C &=& \cos \left(\frac{1}{3} \cos ^{-1}\left(\frac{3 \sqrt{3} \left(1-\lambda ^2\right) \left(s_1^2+s_2^2+1\right)}{\left(3-\lambda ^2+2s_1^2 + 2 s_2^2 + 4 s_1^2 s_2^2 \right)^{3/2}}\right)\right),
\eea
Following the discussion in section \ref{sec:photon}, we find that the size factors are given by
\bea
R_{\rm ph}^2 &=& a^2+4 m^2 s_1^2 s_2^2+2 m \left(s_1^2+s_2^2\right) r_{\rm ph}+r_{\rm ph}^2\,,\nn\\
R_{\rm sh}^2 &=& \fft{2}{(r_{\rm ph}-m)^2}\Big(r_{\rm ph}^4 + 4 m \left(s_1^2+s_2^2\right)r_{\rm ph}^3\nn\\
&&+\left(a^2+m^2 \left(2 s_1^4+8 s_2^2 s_1^2-6 s_1^2+2 s_2^4-6 s_2^2-3\right)\right)r_{\rm ph}^2\nn\\
&&+2m\left(a^2 \left(s_1^2+s_2^2\right)+m^2 \left(-2 s_1^4-8 s_2^2 s_1^2-s_1^2-2 s_2^4-s_2^2\right)\right)r_{\rm ph}\nn\\
&&+m^2\left(a^2 \left(2 s_1^4+4 s_2^2 s_1^2+2 s_1^2+2 s_2^4+2 s_2^2+1\right)+4 m^2 s_1^2 s_2^2\right)
\Big).
\eea
We now find that
\bea
{\cal X}^{-2} &=& \fft{1}{27 \gamma^2 C^2 \left(s_1^2+s_2^2+1\right)^2}\Big(12 \gamma ^3 C \left(s_1^2+s_2^2+1\right)\nn\\
&&
+\gamma ^2 \left(4 C^2 \left(s_1^4+\left(6 s_2^2+4\right) s_1^2+s_2^4+4 s_2^2+3\right)+9 \left(s_1^2+s_2^2+1\right)^2\right)\nn\\
&&-6 \left(s_1^2+s_2^2+1\right)^2 \left(\left(2 s_2^2+1\right) s_1^2+s_2^2+1\right)\Big),\cr
{\cal Y}^2&=&\fft{1}{3 \gamma ^2 C^2 \left(4 \gamma ^2 C^2-3 \gamma^2+4 \gamma  C \left(s_1^2+s_2^2+1\right)+4 \left(\left(2 s_2^2+1\right) s_1^2+s_2^2+1\right)\right)}\Big(\nn\\
&&4 \gamma ^2 C^2 \left(s_1^4+\left(6 s_2^2+4\right) s_1^2+s_2^4+4 s_2^2+3\right)+12 \gamma^3 C \left(s_1^2+s_2^2+1\right)\nn\\
&&+9 \gamma ^2 \left(s_1^2+s_2^2+1\right)^2-6 \left(s_1^2+s_2^2+1\right)^2 \left(\left(2 s_2^2+1\right) s_1^2+s_2^2+1\right)\Big),\nn\\
{\cal Z}^2&=&\fft{6 \left(-3 \gamma ^2+4 \gamma ^2 C^2+4 \gamma  C+4 s_2^2 (\gamma  C+1)+4 s_1^2 \left(\gamma  C+2 s_2^2+1\right)+4\right)}{27\left(\sqrt{1-\lambda ^2} \left(s_1^2+s_2^2+1\right)+\left(2 s_2^2+1\right) s_1^2+s_2^2+1\right)}\,.
\eea
Note that we give ${\cal X}^{-2}$ here instead of ${\cal X}^2$ for the presentation purpose. In general we can perform numerical plots and see that $\cal (X,Y,Z)$ are all no less than 1. In the extremal $\lambda=1$ limit, the expressions are much simpler and we have
\bea
{\cal X}^2 &=& \frac{27/4}{1+\frac{2 \left(\sqrt{2} \sqrt{\left(2 s_2^2+1\right) s_1^2+s_2^2+1} \left(s_1^2+s_2^2+1\right)+\left(2 s_2^2+1\right) \left(s_1^2+s_2^2+1\right)-2 \left(s_2^4+s_2^2\right)\right)}{\left(s_1^2+s_2^2+1\right)^2}},\nn\\
{\cal Y}^2 &=& \frac{\sqrt{2} s_1^2+2 \sqrt{\left(2 s_2^2+1\right) s_1^2+s_2^2+1}+\sqrt{2} \left(s_2^2+1\right)}{3 \sqrt{\left(2 s_2^2+1\right) s_1^2+s_2^2+1}},\nn\\
{\cal Z}^2&=& \frac{4 \sqrt{2} s_1^2+8 \sqrt{\left(2 s_2^2+1\right) s_1^2+s_2^2+1}+4 \sqrt{2} \left(s_2^2+1\right)}{9 \sqrt{\left(2 s_2^2+1\right) s_1^2+s_2^2+1}}.
\eea
Keeping in mind that both $s_1^2$ and $s_2^2$ run from 0 to infinity, it is easy to demonstrate analytically that these quantities are no less than 1.

For $\theta_0\ne 0$, the relevant $r_{\rm ph}^\pm$ are determined by the sextic polynomial associated with
$y(r_{\rm rp}^\pm)=0$.  We do not have a clever procedure; instead we can randomly choose specific numerical numbers for the parameters $(m,a\le m, s_1,s_2)$ and then numerically solve for $r_{\rm ph}^\pm$ from $y=0$. This allows us to evaluate the associated $R_{\rm ph}$ and $R_{\rm sh}$ and validate the conjecture (\ref{conjecture}).  We find no counter example with this incomplete numerical approach.

In the extremal limit $m=a$, we find that $r_{\rm ph}^\pm$ can be solved exactly for $\theta_0=\pi/2$, given by
\bea
r_{\rm ph}^+ &=& 2m(1 + c_1 c_2)\,,\nn\\
r_{\rm ph}^- &=&
\left\{
  \begin{array}{ll}
    2m s_1 s_2, &\qquad 2m s_1s_2\ge a\,, \\
    a, &\qquad 2m s_2 s_2\le a\,.
  \end{array}
\right.
\eea
When $2ms_1 s_2\ge a$, we have
\bea
{\cal X}&=& \frac{3 \sqrt{3} \left(s_1^2+s_2^2+1\right)}{2 \left(c_1 c_2+s_1^2+s_2 s_1+s_2^2+1\right)}\,,\nn\\
{\cal Y}&=& \frac{2 \left(c_1 c_2+s_1^2+s_2 s_1+s_2^2+1\right)}{\sqrt{3c_1 c_2} \left(c_1+c_2\right)+\sqrt{3s_1 s_2} \left(s_1+s_2\right)}\,,\nn\\
{\cal Z}&=& \frac{\sqrt{2c_1 c_2} \left(c_1+c_2\right)+\sqrt{2s_1 s_2} \left(s_1+s_2\right)}{3 \sqrt{2 s_2^2 s_1^2+s_1^2+s_2^2+1}}\,.
\eea
When $2ms_1 s_2\le  a$, we have
\bea
{\cal X}&=& \frac{6 \sqrt{3} \left(s_1^2+s_2^2+1\right)}{4 c_1 c_2+4 \left(s_2^2+1\right) s_1^2+4 s_2^2+5}\,,\nn\\
{\cal Y}&=& \frac{4 c_1 c_2+4 \left(s_2^2+1\right) s_1^2+4 s_2^2+5}{2 \sqrt{3c_1 c_2} \left(c_1+c_2\right)+\sqrt{3\left(2 s_1^2+1\right) \left(2 s_2^2+1\right)}}\,,\nn\\
{\cal Z}&=& \frac{2 \sqrt{c_1 c_2} \left(c_1+c_2\right)+\sqrt{\left(2 s_1^2+1\right) \left(2 s_2^2+1\right)}}{3 \sqrt{2(1 + s_1^2 + s_2^2 + 2 s_1^2 s_2^2)}}\,.
\eea
The above two sets of quantities become the same when $2ms_1 s_2=a$.  In this extremal case, it is not hard to demonstrate analytically that $\cal (X,Y,Z)$ are all greater than 1.

\subsection{General charges}

The situation becomes more much complicated for the four generic charges.  We find that the photon surfaces are determined by
\bea
\xi &=& \fft{1}{a^2(m-r)}\Big[am \big(c_{1234} \left(a^2-r^2\right)+s_{1234} \left((r-2 m)^2-a^2\right)\big)\nn\\
 &&+ a \Delta\, \big(r^2 + m (r-m) \sum_{i} s_i^2 + m^2 ((c_{1234} - s_{1234})^2-1)\big)^{\fft12}
\Big]_{r_{\rm ph}},\cr
\eta &=&\fft{1}{a^2(m-r)^2}\Bigg[-r^6 + m r^5 \sum_i c_i^2 - m^2 r^4 \Big(2 (c_{1234}-s_{1234})^2 -3 -5
\sum_i s_i^2\Big)\nn\\
&&-mr^3\Big(4m^2 \big((c_{1234} - 3s_{1234})(c_{1234}-s_{1234})+ 1 + 2\sum_i s_i^2\big)\nn\\
&&\qquad\quad - a^2 (2 + \sum_i s_i^2)\Big)\nn\\
&&-m^2 r^2\Big( 2(4m^2 + a^2) (c_{1234} - s_{1234})^2 -4m^2(c_{1234}^2 - 5 s_{1234}^2+1 +\sum_i s_i^2)\nn\\
&&-a^2\big(2 + 3\sum_i s_i^2 + 4 \sum_{i<j} s_{ij}^2\big)\Big)\cr
&&+ 8m^3 r\Big(a^2 \big(c_{1234} (c_{1234}-s_{1234})+ 1 + \sum_i s_i^2 + \sum_{i<j} s_{ij}^2\big) +4m^2 s_{1234}^2\Big)\nn\\
&&-4m^4 \Big(4m^2 s_{1234}^2 + a^2 \sum_{i<j<k} s_{ijk}^2\Big)+ 2m \Delta\, \big(c_{1234} - (2m-r)^2 s_{1234}\big)\nn\\
&&\qquad \times \big(r^2 + m (r-m) \sum_{i} s_i^2 + m^2 ((c_{1234} - s_{1234})^2-1)\big)^{\fft12}\Bigg]_{r_{\rm ph}}.
\eea
We have to employ numerical technique to determine the size factors $R_{\rm ph}$ and $R_{\rm sh}$, even though the procedure was well specified in section \ref{sec:photon}.  An exhaustive analysis of all parameters is beyond our current numerical skill; however, for a large number of randomly selected mass and charges, we have found no counterexample to our conjecture.

\section{Conclusions}

In this paper, we studied the photon surfaces and shadows for asymptotically flat rotating black holes in four dimensions. Our key motivation was to characterize the size of these geometries.  For spherically symmetric black holes, the radii are the noncontroversial parameters, but the choice is non-unique when the black hole is rotating, since the shapes of the shadow can be distorted by rotations and they are also dependent on the angular position of the observer.

We used the area for measuring the size of the horizon, but continued to use linear length to measure the size of both shadow and the relevant photon orbits. The reason for the former is the curious inequality (\ref{sGthetatheta}). The reason for the latter is because there exists a shortest diagonal line across a shadow and therefore it is natural to use its half length as the measure of the shadow.  Our choice of the parameter measuring the size of the photon orbits may appear to {\it ad hoc}, but we believe that our choice captures the relevant photon orbits that is responsible to create the boundary of the shadow.

After having determined the parameters, we then verified the conjecture (\ref{conjecture}), established for static black holes in \cite{Lu:2019zxb}, with a variety of rotating black holes. These are the Kerr, Kerr-Newman, Kerr-Sen and Kerr-Cveti\v c-Youm black holes. For some examples, we could validate the conjecture analytically or by numerical plots for all range of parameters.  For more complicated examples, we checked a large number of data, {\it albeit} incomplete, and we found no counterexample.

The number of known rotating black holes is much less than static ones; nevertheless, the success of our verification does validate our choice of the size parameters and it also indicates some deep underlying principle behind the conjecture (\ref{conjecture}).  Nevertheless, the procedure depends heavily on the coordinate choice, despite of our effort to remove the ambiguity with our strict gauge choice. It is thus of great interest to examine the conjecture (\ref{conjecture}) using the areas, not only of the horizon, but also the shadow.  Our preliminary investigation in appendix \ref{sec:app-shadow} indicates that the shadow areas ${\cal A}_{\rm sh}$ are also consistent with the conjecture (\ref{conjecture}). To be precise, we expect that the shadow areas of black holes in Einstein gravity satisfy both the lower and upper bounds
\be
\fft{S}{4\pi}\, \le\, \fft{{\cal A}_{\rm sh}}{27\pi}\,\le\, M^2\,.\label{areaconjecture}
\ee
The Schwarzschild black hole saturates both.

\section*{Acknowledgement}

We are grateful to Li-Ming Cao, Zhao-Long Wang, Jun-Bao Wu and Run-Qiu Yang for useful discussions. H.L.~is grateful to Institute of Modern Physics, Northwest University for hospitality during the early stage of this work. This work is supported in part by NSFC (National Natural Science Foundation of China) Grant No.~11935009.  X.H.~Feng is also supported in part by NSFC Grant No.~11905157. H.L.~is also supported by NSFC grant No.~11875200.

\appendix

\section{Shadows of extremal Kerr black hole}
\label{sec:app-shadow}

In this appendix, we present a few numerical plots of the black hole shadows, which can help to illustrate our definition of size factors discussed in both introduction and section \ref{sec:photon}.  The shapes of the shadows are analogous for various black holes, we therefore use the Kerr black hole as an illustrative example.  The purpose of drawing these shadows is to help us to characterize their size.

\begin{figure}[ht!]
\begin{center}
\includegraphics[width=120pt]{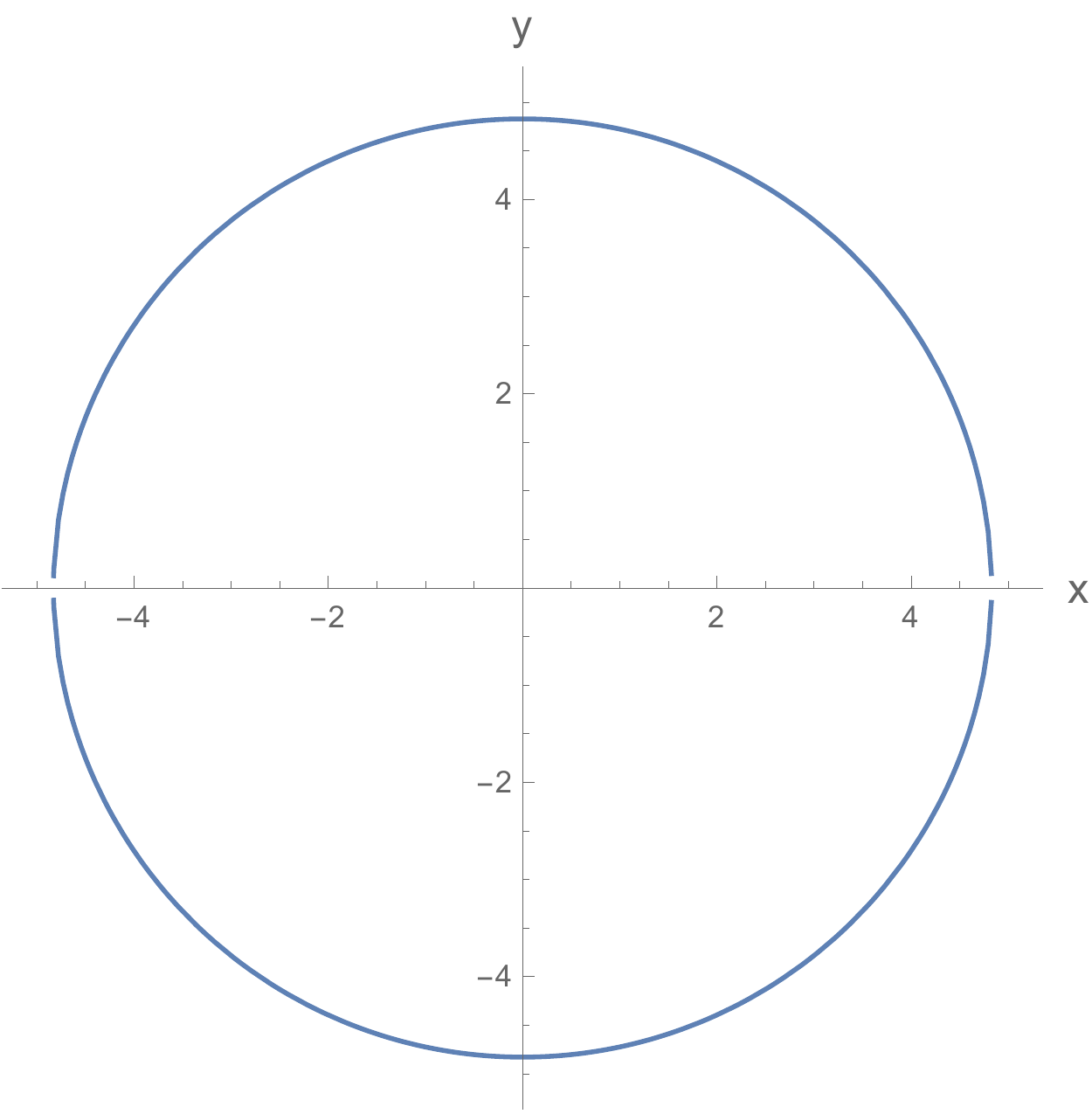}
\includegraphics[width=120pt]{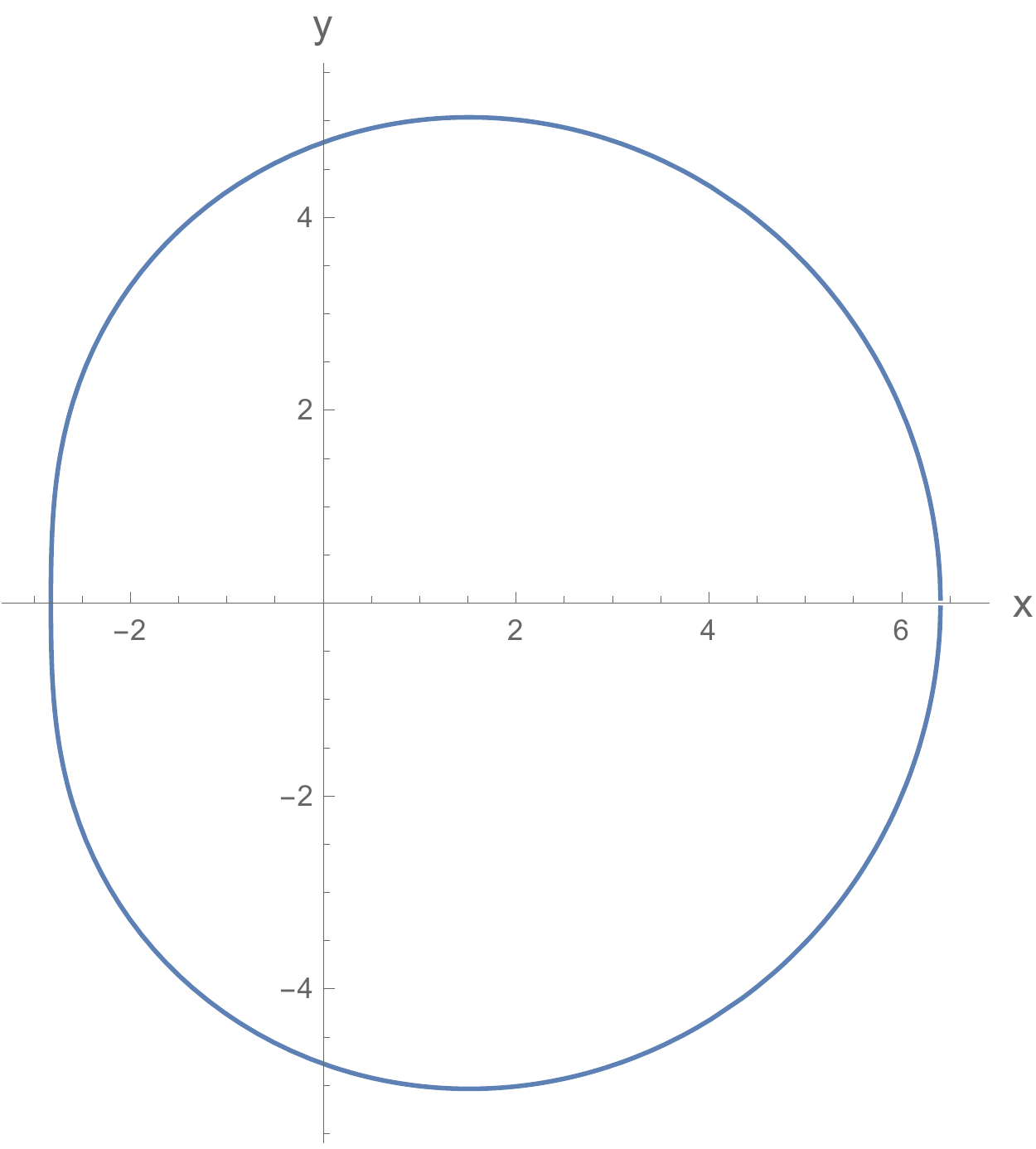}
\includegraphics[width=120pt]{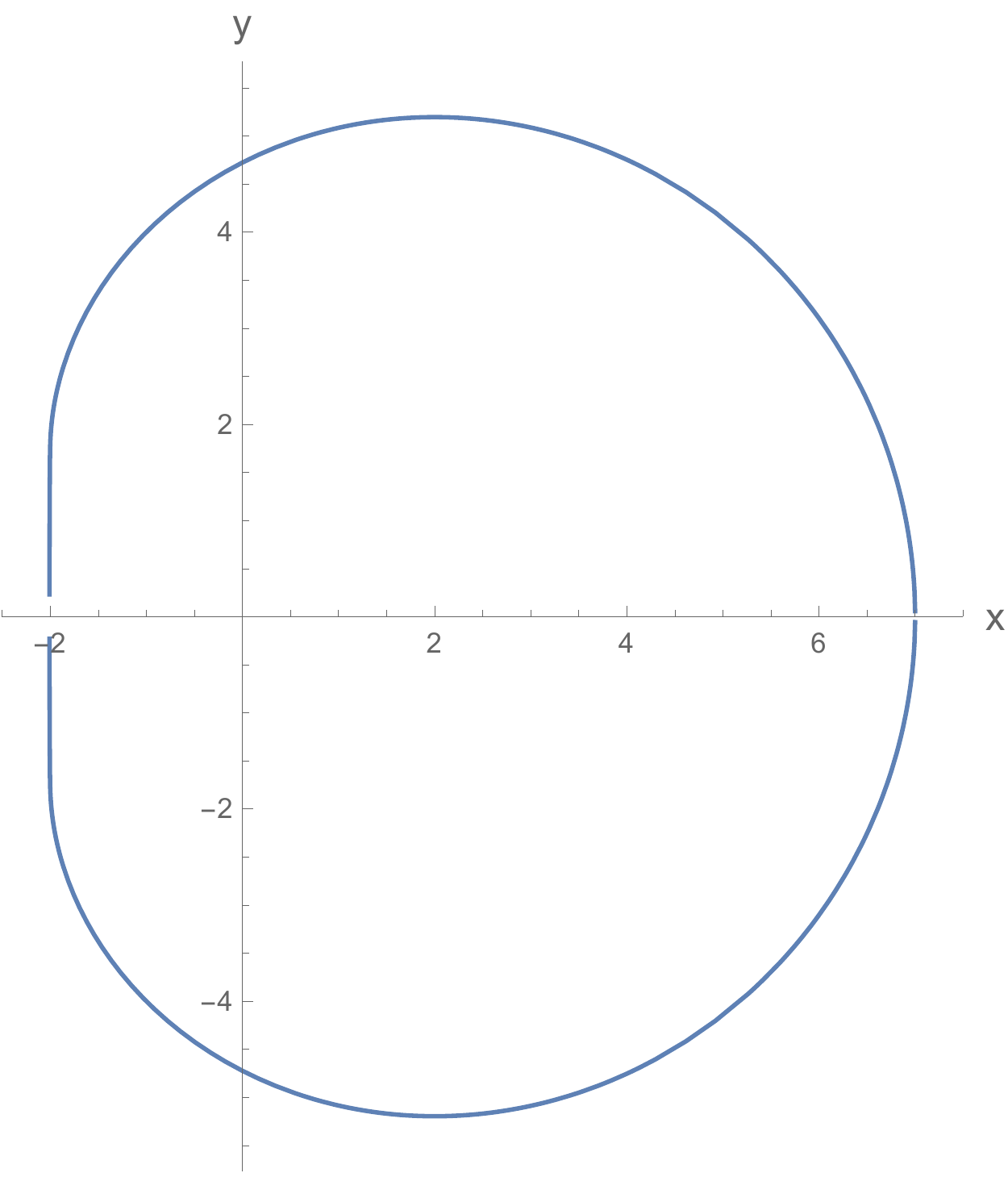}
\end{center}
\caption{{\it Shadows of the extremal Kerr black hole of $m=1=a$, corresponding to $M=1$ and $R_+=\sqrt{2}$. The left one is the shadow viewed from the north pole with $r_{\rm ph}=1+\sqrt2$ and hence $R_{\rm ph}=\sqrt{2(2+\sqrt2)}$ and $R_{\rm sh}=2(1 + \sqrt2)$.
The middle plot is viewed from angle $\theta_0=\pi/4$, we have $r_{\rm ph}^-=1.0583$ and $r_{\rm ph}^+=3.5549$, giving $x^-=-2.824$ and $x^+=6.403$, and hence we have $R_{\rm ph}=2.449$ and $R_{\rm sh}=4.613$. The right plot is the shadow viewed from $\theta_0=\pi/2$ and we have $x^-=- 2 $ at $r_{\rm ph}^-=1$ and $x^+= 7$ at $r_{\rm ph}^-=4$; therefore, we have $R_{\rm ph}=5/2$ and $R_{\rm sh}=9/2$. In all the three cases, we have ${\cal X}>1$,${\cal Y}>1$ and ${\cal Z}>1$, for $\cal (X,Y,Z)$ defined in (\ref{XYZ}). As a comparison, for the $M=1$ Schwarzschild black hole, we have $R_+=2$, $R_{\rm ph}=3$ and $R_{\rm sh}=3\sqrt3$. Thus rotation makes the black hole to appear ``smaller''.}}
\label{fig1}
\end{figure}

In Fig.~\ref{fig1}, we present the shape of the shadows of the extremal Kerr black hole viewed from the north pole $\theta_0=0$, $\theta=\pi/4$ and the equatorial plane $\theta_0=\pi/2$.  From the north pole, the shadow is round, but smaller than that of the Schwarzschild black hole of equal mass.  The image becomes most skewed when $\theta_0=\pi/2$. To understand the skewing effect, we note that for our coordinate convention, the angular momentum vector pointing to the north pole direction for positive $a$.  It follows from the geodesic equations that, for the two orbits $r_{\rm ph}^+> r_{\rm ph}^-$, we have
\bea
&& \fft{d\phi}{d\tau} < 0\,,\qquad r_{\rm ph} = r_{\rm ph}^+\,,\nn\\
&& \fft{d\phi}{d\tau} > 0\,,\qquad r_{\rm ph} = r_{\rm ph}^-\,.
\eea
In other words, the photon orbiting in the opposite direction of the black hole rotation has bigger $r_{\rm ph}^+$ and bigger $|x^+|$.

For all the cases, the inequalities we propose in this paper are all satisfied, with \be
\{{\cal X,Y,Z}\}=
\left\{
  \begin{array}{ll}
    \{1.0762, 1.0668, 1.2318\}, &\qquad \theta_0=0 \\
    \{1.1264, 1.0877, 1.1543\}, &\qquad \theta_0=\fft{\pi}4 \\
    \{1.1547, 1.0392, 1.1785\}, &\qquad \theta_0=\fft{\pi}2
  \end{array}
\right.
\ee
for the left, middle and right plots respectively.

It is also instructive to calculate the area of the three shadows presented in Fig.~\ref{fig1}.  The general formula is given by
\be
{\cal A} = \int_{r_{\rm ph}^-}^{r^+_{\rm ph}} dr_{\rm ph}\, y x'\,.
\ee
where a prime denotes a derivative with respect to $r_{\rm ph}$.  We find that for the extremal Kerr black hole with $m=a=1$, the shadow areas are given by
\bea
\theta_0 =0:&& \fft{{\cal A}_{\rm sh}}{27\pi} = \ft{2}{27} (\sqrt{2}+2)^2\sim 0.8635\,,\nn\\
\theta_1 =\ft14\pi:&& \fft{{\cal A}_{\rm sh}}{27\pi} \sim 0.8766\,,\nn\\
\theta_2 = \ft12 \pi:&&\fft{{\cal A}_{\rm sh}}{27\pi} =\ft{1}{27}(15 \sqrt{3}+16 \pi) \sim 0.8989\,.
\eea
The results indicate that the area of the photon shadow is a monotonically increasing function of $\theta_0$.
If we define the mean radius of the shadow radius by $R_{\rm sh} = \sqrt{{\cal A}_{\rm sh}/\pi}$, then the values are also consistent with our main conjecture (\ref{conjecture}), yielding the area conjecture (\ref{areaconjecture}).

\begin{figure}[ht!]
\begin{center}
\includegraphics[width=120pt]{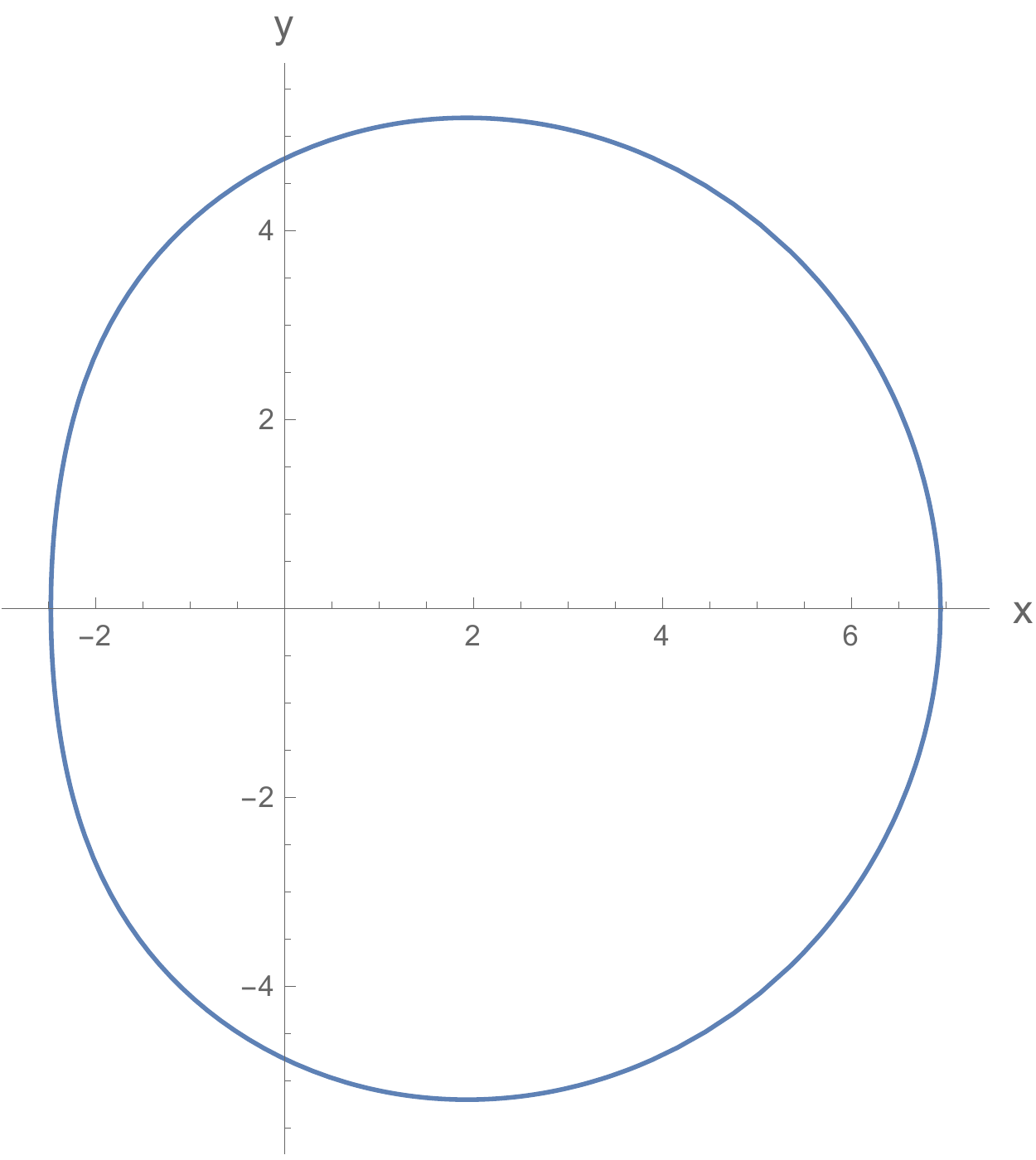}
\end{center}
\caption{{\it The shadow of non-extremal Kerr black hole with $m=1$ and $a=29/30$, viewed from the equatorial plane ($\theta_0=\pi/2$). Note the similarity between this and the middle plot of Fig.~\ref{fig1} which is the shadow of the extremal black hole but viewed from angle $\theta_0=\pi/4$}.}
\label{fig2}
\end{figure}

 We also present the shadow plot in Fig.~\ref{fig2} for the non-extremal Kerr black hole with mass $m=1$ and $a=19/20$, viewed from the equatorial plane. There is a strong similarity between this shadow and the one of the extremal black hole viewed from the angle $\theta_0=\pi/4$, shown in the middle plot of Fig.~\ref{fig1}. It is thus difficult to distinguish these situations without precision measurement.  Nevertheless, our conjecture (\ref{conjecture}) appears to be valid in all these cases.

\section{Roots of quartic equation}
\label{sec:quartic}

In a few of our black hole examples, we are required to solve the quartic polynomial equation
\be
a r^4+b r^3+c r^2+dr +e=0\,,\qquad \hbox{with}\qquad a\ne0\,,
\ee
where $(a,b,c,d,e)$ are constants specified by the black hole mass, angular momentum and charges. We follow the wikipedia entry and the general solutions is
\bea
x_{1,2}=-\frac{b}{4a}-S\pm\frac{1}{2}\sqrt{-4S^2-2p+\frac{q}{S}}\,,&&\\
x_{3,4}=-\frac{b}{4a}+S\pm\frac{1}{2}\sqrt{-4S^2-2p-\frac{q}{S}}\,,
\eea
where
\bea
p&=&\frac{8ac-3b^2}{8a^2}\,,\qquad q=\frac{b^3-4abc+8a^2d}{8a^3}\,,\nn\\
S&=&\frac{1}{2}\sqrt{-\frac{2}{3}p+\frac{1}{3a}\left(Q+\frac{\Delta_0}{Q}\right)}\,,\qquad
Q=\sqrt[3]{\fft12 \left({\Delta_1+\sqrt{\Delta_1^2-4\Delta_0^3}}\right)}\,,\nn\\
\Delta_0&=&c^2-3bd+12ae\,,\qquad
\Delta_1= 2c^3-9bcd+27b^2e+27ad^2-72ace\,.
\eea
For charged rotating black holes, we have either two real roots and two conjugate complex roots or four real roots. We require that the photon sphere is located outside of the horizon, which implies that
\be
r_{\rm ph}^{\pm}=-\frac{b}{4a}+S\pm\frac{1}{2}\sqrt{-4S^2-2p-\frac{q}{S}}\,.
\ee

\end{document}